\documentclass{decar-wsd}    
\usepackage{decar-common}    
\usepackage{decar-dynamics}  

\usepackage{amsmath}
\usepackage{graphicx}
\usepackage{float}
\usepackage{caption}
\usepackage{lipsum}
\usepackage[hidelinks]{hyperref}

\title{Asymptotically Stable Data-Driven Koopman Operator Approximation with Inputs using Total Extended DMD}
\author{Louis Lortie and James Richard Forbes}
 
\begin{document}

\makeatletter
\def\@maketitle{%
  \null
  \vskip 2em%
  \begin{center}%
  \let \footnote \thanks
    {\LARGE \textbf{\@title} \par}%
    \vskip 1.5em%
    {\large
      \lineskip .5em%
      \begin{tabular}[t]{c}%
        \emph{\@author} \\
        \small{Department of Mechanical Engineering, McGill University} \\
        \small{817 Sherbrooke Street West, Montreal QC H3A 0C3} \\
        \small{E-mail: louis.lortie@mail.mcgill.ca}
      \end{tabular}\par}%
    \vskip 1em%
  \end{center}%
  \par
  \begin{center}%
    \rule{\linewidth}{1.5pt}%
  \end{center}%
  \raggedright}
\makeatother

\maketitle


\begin{abstract}
The Koopman operator framework can be used to identify a data-driven model of a nonlinear system. Unfortunately, when the data is corrupted by noise, the identified model can be biased. Additionally, depending on the choice of lifting functions, the identified model can be unstable, even when the underlying system is asymptotically stable. This paper presents an approach to reduce the bias in an approximate Koopman model, and simultaneously ensure asymptotic stability, when using noisy data. Additionally, the proposed data-driven modeling approach is applicable to systems with inputs, such as a known forcing function or a control input. Specifically, bias is reduced by using a total least-squares, modified to accommodate inputs in addition to lifted inputs. To enforce asymptotic stability of the approximate Koopman model, linear matrix inequality constraints are augmented to the identification problem. The performance of the proposed method is then compared to the well-known extended dynamic mode decomposition method and to the newly introduced forward-backward extended dynamic mode decomposition method using a simulated Duffing oscillator dataset and experimental soft robot arm dataset.
\end{abstract}

\section{Introduction}
Dynamic models of physical systems are used for many tasks, such as prediction, sensitivity analysis to initial conditions or parameter variations, and design. For complex systems, data-driven approaches to modeling \cite{Billings2013, Kutz2016, Brunton2021, Mauroy2020, Brunton2022, Kutz2013} are an attractive alternative to first-principles modeling, since a model structure can be assumed and then data can be used to fit model parameters. The Koopman operator~\cite{Mezic2013, Williams2015, Mauroy2020, Korda2018} is becoming a popular modeling approach owing to its ability to represent a nonlinear system as an infinite-dimensional linear system. When working directly with data, the Koopman operator can be approximated, leading to a finite-dimensional approximate Koopman model~\cite{Mezic2020, Budivsic2012}. The literature provides different methods to compute the finite-dimensional Koopman representation of autonomous and non-autonomous systems, such as with neural networks \cite{Xiong2024, Liu2024} and via least squares \cite{Kutz2016, Hemati2017}. The lifting functions associated with the approximate Koopman model must be able to represent a rich enough set of nonlinearities in order to accurately describe the nonlinear behavior of the system. Depending on whether there are many lifting functions or many snapshots of the data, the Koopman operator is approximated using dynamic mode decomposition (DMD)~\cite{Schmid2010, Proctor2016} or using extended DMD (EDMD)~\cite{Williams2015}, respectively. An approximate Koopman model can then easily be represented as a state-space system \cite{Kutz2016}, which can be used for other tasks, such as prediction \cite{Brunton2021, Korda2020}, stability analysis \cite{Mauroy2016}, or control design \cite{Bruder2019, Bruder2020, Korda2018}.

Identifying a real system from data involves measurements, and measurements are always corrupted by some amount of noise. When a dynamic model is fit using noisy data, the data-driven dynamic model can be biased~\cite{Dawson2016, Hemati2017}. The bias in the Koopman operator approximation is reflected in the dynamics and input matrices. The bias can impact the eigenvalues of the dynamics matrix, resulting in eigenvalues that are shifted towards the origin of the complex plan, resulting in higher than expected decay rates~\cite{Dawson2016}. Many papers in the literature provide methods to reduce the bias in the dynamics matrix, such as forward-backward DMD (fbDMD)~\cite{Dawson2016} and total DMD (TDMD)~\cite{Golub1980, Hemati2017}. Recent work \cite{Lortie2024} has adapted fbDMD to also account for the bias in the input matrix, but there is no equivalent adaptation for the input matrix using TDMD. Extending TDMD to reduce the bias in the input matrix has the potential to outperform \cite{Lortie2024}, since the authors in \cite{Dawson2016} show that TDMD reduces the bias more than fbDMD in the dynamics matrix. Additionally, when identifying an inherently asymptotically stable system, the resulting model can be unstable if the lifting functions are chosen poorly. If the underlying dynamics are asymptotically stable, then the data-driven model must be asymptotically stable as well, else the data-driven model is not at all representative of the true system and is not useful for tasks such as prediction. Some work from the literature provides constraints formulated as linear matrix inequality (LMI) in the Koopman operator approximation problem to enforce asymptotic stability in the Koopman least-squares problem \cite{Dahdah2022} and in the Koopman fbEDMD problem \cite{Lortie2024}.

TDMD \cite{Hemati2017}, inspired by total least-squares DMD \cite{Golub1980, Markovsky2007}, projects the snapshot matrices onto an augmented matrix to reduce the bias in the dynamics matrix. This method is different than classic least squares, since it minimizes the orthogonal distance between the linear fit and the data points, while least squares minimizes the vertical distance between the linear fit and the data points \cite{Hemati2017}. This paper proposes a method to extend the application of TDMD to include both the dynamics and the input matrices associated with the approximate Koopman model. Including inputs to the TDMD framework allows for the consideration of a wider range of applications, such as regulated systems and engineering applications requiring inputs. Additionally, to ensure that the proposed method identifies an asymptotically stable Koopman system with reduced bias, this paper introduces a new formulation of the TDMD problem using LMI constraints to enforce asymptotic stability. In summary, the proposed method is a two-part method, which firstly projects the snapshot matrices with inputs onto an augmented matrix, and secondly computes the Koopman matrix with reduced bias by solving a convex optimization problem that imposes asymptotic stability on the identified system. The performance of the proposed method is compared to the state-of-the-art methods, forward-backward EDMD (fbEDMD) \cite{Lortie2024} and EDMD \cite{Williams2015}, using a simulated dataset of a Duffing oscillator and an experimental dataset of a soft robot arm. 

\subsection{Related work}

The stochastic nature of sensor noise creates difficulties when employing system identification methods with noisy data. Some work in the literature proposes methods to mitigate the impact of noise on the dynamics matrix~\cite{Haseli2019, Hemati2017, Dawson2016}. In~\cite{Dawson2016}, fbDMD is introduced as a method to leverage the forward- and backward-in-time dynamics of a system to reduce the bias in the eigenvalues of the dynamics matrix. Expanding on \cite{Dawson2016}, the authors in \cite{Lortie2024} introduce fbEDMD to also reduce the bias in the input matrix. An additional method to reduce bias in the dynamics matrix of the Koopman system is TDMD \cite{Hemati2017}. TDMD is a two-step procedure in which 1) the snapshot matrices are projected on an augmented snapshot matrix, and then 2) the Koopman operator is approximated by solving a least-squares problem with the projected snapshot matrices. Physics-informed DMD (piDMD) \cite{Baddoo2023} is another method to reduce the impact of noise in the identification process of the Koopman operator approximation. In particular, piDMD enforces properties on the solution, such as conservation of energy, self-adjointness, causality, and shift-equivariance. Sensor noise can be characterized by various distributions, such as a Gaussian distribution, but other types of disturbances such as outliers in the data can be complex to address. In \cite{Abolmasoumi2022}, outlier rejection is performed by solving for weights based on a loss function.

An inherently asymptotically stable system can be identified as an unstable Koopman system if the lifting functions are chosen poorly. Additionally, in some cases, noise in the data can also result in the identification of an unstable Koopman system, even when the underlying system is asymptotically stable~\cite{Mamakoukas2020}. There are methods, such as naturally structured DMD (NSDMD), which preserve the stability properties of the original system \cite{Huang2018}. However, NSDMD is applicable to unforced systems and data not corrupted by noise. In order for the dynamic model to be as representative as possible, system identification methods should enforce asymptotic stability when required. In \cite{Dahdah2021, Dahdah2022}, the authors enforce asymptotic stability on the approximate Koopman system with inputs by formulating the Koopman approximation problem as a series of LMI and bilinear matrix inequality (BMI) constraints. To formulate the Koopman operator approximation problem as a convex optimization problem, the BMI constraints used to enforce asymptotic stability on the Koopman system are transformed into a set of LMI constraints in \cite{Lortie2024} by using the method proposed in \cite{Mabrok2023, Lortie2024, Hara2020, Hara2021}. In this paper, the Koopman operator approximation problem is formulated as a series of LMIs to enforce asymptotic stability leveraging the approach taken in \cite{Dahdah2022, Lortie2024}. 

\subsection{Contributions}
This paper presents a new approximate Koopman modeling method, based on TDMD, that 1) reduces the bias in the dynamics and input matrices and 2) enforces asymptotic stability on the approximate Koopman system. The goal of this method is to identify an asymptotically stable Koopman representation with reduced bias when using noisy data regardless of the choice of lifting functions. 

This paper derives the proposed method, \emph{total EDMD with inputs and asymptotic stability constraint}, and validates it against EDMD \cite{Williams2015}, EDMD with an asymptotic stability constraint \cite{Dahdah2021}, fbEDMD \cite{Lortie2024}, fbEDMD with an asymptotic stability constraint \cite{Lortie2024}, and total EDMD with inputs using a simulated dataset of a Duffing oscillator and an experimental dataset of a soft robot arm. 

\subsection{Organization}
This paper is organized in the following way. A review of pertinent background theory relevant to the proposed method is presented in Section 2. The proposed method is derived in Section 3 and formulated as a convex optimization problem in Section 4. Results using simulated and experimental datasets are presented in Section 5, and concluding remarks are given in Section 6.

\section{Background theory} \label{background}

\subsection{Koopman operator theory}
Consider a system whose discrete-time dynamics are governed by the nonlinear difference equation
\begin{equation} \label{disc_time_system}
    \mbf{x}_{k+1} = \mbf{f}(\mbf{x}_k, \mbf{u}_k),
\end{equation}
where $\mbf{x}_k \in \mathcal{N} \subseteq \mathbb{R}^{n \times 1}$, ${\mbf{u}_k \in \mathcal{M} \subseteq \mathbb{R}^{m \times 1}}$, and $\mbf{f}: \mathbb{R}^{n \times 1} \times \mathbb{R}^{m \times 1} \rightarrow \mathbb{R}^{n \times 1}$. The Koopman operator ${\mathcal{U}: \mathcal{H} \rightarrow \mathcal{H}}$,  which operates in the Hilbert space $\mathcal{H}$ of infinitely many lifting functions ${\psi : \mathcal{N} \times \mathcal{M} \rightarrow \mathbb{R}}$, advance such lifting functions forward in time \cite{Kutz2016, Dahdah2022} as  
\begin{equation}
    (\mathcal{U}\psi)(\mbf{x}_k, \mbf{u}_k) = \psi(\mbf{x}_{k+1}, \ast),
\end{equation}
where $\ast = \mbf{u}_k$ or $\ast = \textbf{0}$ if inputs of the system depend on the states of the system or if they do not, respectively. In practice, the Koopman operator is approximated using a finite number of lifting functions partitioned as
\begin{equation}
    \mbs{\psi}(\mbf{x}_k, \mbf{u}_k) = \begin{bmatrix}
        \mbs{\vartheta}(\mbf{x}_k) \\
        \mbs{\upsilon}(\mbf{x}_k, \mbf{u}_k)
    \end{bmatrix},
\end{equation}
where $\mbs{\psi} : \mathcal{M} \times \mathcal{N} \rightarrow \mathbb{R}^{p \times 1}$, $\mbs{\vartheta} : \mathcal{M} \rightarrow \mathbb{R}^{p_\vartheta \times 1}$, $\mbs{\upsilon} : \mathcal{M} \times \mathcal{N} \rightarrow \mathbb{R}^{p_\upsilon \times 1}$, and $p = p_\vartheta + p_\upsilon$. The approximated Koopman operator then advances the finite number of lifting functions as
\begin{equation} \label{disc_time_koop}
    \mbs{\vartheta}(\mbf{x}_{k+1}) = \mbf{U}\mbs{\psi}(\mbf{x}_k, \mbf{u}_k) + \mbf{r}_k,
\end{equation}
where the residual $\mbf{r}_k$ exists due to the Koopman operator being approximated. The Koopman matrix is defined as
\begin{equation} \label{koop}
    \mbf{U} = \begin{bmatrix}
        \mbf{A} & \mbf{B}
    \end{bmatrix},
\end{equation}
which is an approximation of the Koopman operator in finite dimensions, and can be substituted into \eqref{disc_time_koop} to obtain the state-space representation
\begin{equation}
    \mbs{\vartheta}(\mbf{x}_{k+1}) = \mbf{A}\mbs{\vartheta}(\mbf{x}_k) + \mbf{B}\mbs{\upsilon}(\mbf{x}_k, \mbf{u}_k) + \mbf{r}_k, \label{lin_eq_og}
\end{equation}
which evolves in the lifted space. 

When the vector-valued lifting function $\mbs{\upsilon}(\cdot)$ acts only on the inputs $\mbf{u}_k$, then the dynamics matrix $\mbf{A}$ and input matrix $\mbf{B}$ respectively describe the evolution of the lifted states and lifted inputs of the system. The dynamics can then be rewritten as
\begin{equation} \label{new_state_eq}
    \mbs{\vartheta}(\mbf{x}_{k+1}) = \mbf{A}\mbs{\vartheta}(\mbf{x}_k) + \mbf{B}\mbs{\upsilon}(\mbf{u}_k) + \mbf{r}_k. 
\end{equation}
The systems of interest in this paper are described by the linear dynamics in \eqref{new_state_eq} and the measurement equation
\begin{equation}\label{output_eq}
    \mbs{\zeta}_k = \mbf{C}\mbs{\vartheta}_k + \mbf{D}\mbs{\upsilon}_k + \mbf{v}_k,
\end{equation}
where, in this paper, $\mbf{v}_k$ represents a vector of white sensor noise, $\mbs{\zeta}_k \in \mathbb{R}^{p_\zeta \times 1}$, $\mbf{C} = \textbf{1}$, and $\mbf{D} = \textbf{0}$.

Let measurements collected from sensors and the inputs to the system over time steps $k = 0, 1, \dots, q$ be lifted and arranged into the snapshot matrices
\begin{equation}
    \mbs{\Psi} = \begin{bmatrix}
        \mbs{\psi}_0 & \mbs{\psi}_1 & \cdots & \mbs{\psi}_{q-1}
    \end{bmatrix} \in \mathbb{R}^{p \times q}, \label{Psi} 
\end{equation}
and
\begin{equation} \label{Theta_+}
    \mbs{\Theta}_+ = \begin{bmatrix}
        \mbs{\vartheta}_1 & \mbs{\vartheta}_2 & \cdots & \mbs{\vartheta}_q 
    \end{bmatrix} \in \mathbb{R}^{p_\vartheta \times q},
\end{equation}
where $\mbs{\psi}_k = \mbs{\psi}(\mbf{x}_k, \mbf{u}_k)$ and $\mbs{\vartheta}_k = \mbs{\vartheta}(\mbf{x}_k)$. Using \eqref{Psi} and \eqref{Theta_+}, the Koopman matrix is computed by solving the least-squares problem
\begin{equation}
    \mbf{U} = \arg \min_{\mbf{U}^\star} \;\;\; \left\|\mbs{\Theta}_+ - \mbf{U}^\star\mbs{\Psi}\right\|^2_\frob, \label{vanilla_least}
\end{equation}
giving the solution~\cite{Kutz2016} 
\begin{equation} \label{true_edmd}
    \mbf{U} = \mbs{\Theta}_+\mbs{\Psi}^\dagger, 
\end{equation}
where $\|\cdot\|_\frob$ is the Frobenius norm. When $q \gg p$, then it is preferred to compute the Koopman matrix using EDMD \cite{Williams2015} in order to reduce the size of the pseudoinverse problem on $\mbs{\Psi}$, such that \eqref{true_edmd} becomes
\begin{equation} \label{forw_edmd1}
    \mbf{U} = \mbs{\Theta}_+\mbs{\Psi}^\dagger = (\mbs{\Theta}_+\mbs{\Psi}^\trans)(\mbs{\Psi}\mbs{\Psi}^\trans)^\dagger = \mbf{G}\mbf{H}^\dagger, 
\end{equation}
where 
\begin{equation} \label{forw_edmd}
    \mbf{G} = \frac{1}{q}\mbs{\Theta}_+\mbs{\Psi}^\trans \in \mathbb{R}^{p_\vartheta \times p}, \hspace{10pt} \mbf{H} = \frac{1}{q}\mbs{\Psi}\mbs{\Psi}^\trans \in \mathbb{R}^{p \times p},
\end{equation}
and $q$ is the number of snapshots.

\subsection{Dynamic mode decomposition}
DMD \cite{Schmid2010} is a data-driven method used to reconstruct the dynamics of a system. In particular, DMD is useful when the system under investigation is complex and high-dimensional in features \cite{Kutz2016, Proctor2016}. Consider the dataset $\mathcal{D} = \{\mbf{x}_k\}^m_{k=0}$ with snapshot matrices 
\begin{equation}
    \mbf{X} = \begin{bmatrix}
        \mbf{x}_0 & \mbf{x}_1 & \cdots & \mbf{x}_{m-1}
    \end{bmatrix} \in \mathbb{R}^{n \times m}, \label{data_states} 
\end{equation}
and
    \begin{equation}
    \mbf{X}_+ = \begin{bmatrix}
        \mbf{x}_1 & \mbf{x}_2 & \cdots & \mbf{x}_{m}
    \end{bmatrix} \in \mathbb{R}^{n \times m}, \label{data_states_+} 
\end{equation}
where $\mbf{x}_k \in \mathbb{R}^{n \times 1}$ is a vector of the system's states.
The subscript $k$ represents the timestep such that $\mbf{x}_k = \mbf{x}(t_k)$ where $k = 0,1,2,\ldots, m$. The best-fit matrix $\mbf{A}$, where
\begin{equation} \label{dynamics_eq}
    \mbf{X}_+ = \mbf{A}\mbf{X}, 
\end{equation}
advances the snapshot matrix in \eqref{data_states} by one time step. The matrix $\mbf{A}$ is the approximation of the dynamics matrix of the system. Solving \eqref{dynamics_eq} is equivalent to solving the least-squares problem
\begin{equation} \label{DMD_least}
    \mbf{A} = \arg \min_{\mbf{A}^\ast} \;\;\; \left\|\mbf{X}_+ - \mbf{A}^\ast\mbf{X}\right\|^2_\frob,
\end{equation}
whose solution is
\begin{equation} \label{DMD}
    \mbf{A} = \mbf{X}_+\mbf{X}^\dagger,
\end{equation}
where $(\cdot)^\dagger$ is the Moore-Penrose pseudoinverse. The objective of DMD is to find the leading eigendecomposition of $\mbf{A}$ when $\mbf{A}$ itself is too computationally demanding to compute or too large to store. Consider the SVD of the snapshot matrix
\begin{equation}
    \mbf{X} = \mbf{W}\mbs{\Sigma}\mbf{V}^\trans,
\end{equation}
where $\mbf{W} \in \mathbb{R}^{n \times n}$, $\mbs{\Sigma} \in \mathbb{R}^{n \times m}$, and $\mbf{V} \in \mathbb{R}^{m \times m}$. It follows that \eqref{DMD} can be written as
\begin{equation} \label{DMD_svd}
    \mbf{A} = \mbf{X}_+\mbf{V}\mbs{\Sigma}^\dagger\mbf{W}^\trans.
\end{equation}
Assume that the dynamics described by \eqref{dynamics_eq} have a low-dimensional structure. Truncating the singular values in \eqref{DMD_svd} captures the main low-dimensional modes~\cite{Kutz2016} while removing unwanted high-dimensional features. In turn, truncating the smaller singular values improves the conditioning of the pseudo-inverse in \eqref{DMD}. After truncation, \eqref{DMD_svd} is rewritten as
\begin{equation} \label{DMD_svd_trun}
    \mbf{A} = \mbf{X}_+\tilde{\mbf{V}}\tilde{\mbs{\Sigma}}^{-1}\tilde{\mbf{W}}^\trans,
\end{equation}
where $\tilde{\mbf{W}} \in \mathbb{R}^{n \times r}$, $\tilde{\mbs{\Sigma}} \in \mathbb{R}^{r \times r}$, $\tilde{\mbf{V}} \in \mathbb{R}^{r \times m}$, and \emph{r} is the number of remaining singular values. When considering the case where $n \gg r$, it is more computationally efficient to compute $\tilde{\mbf{A}}$, the projection of the dynamics matrix $\mbf{A}$ on the left singular vectors $\tilde{\mbf{W}}$ as \cite{Kutz2016, Proctor2016}
\begin{equation}
    \tilde{\mbf{A}} = \tilde{\mbf{W}}^\trans\mbf{A}\tilde{\mbf{W}} = \tilde{\mbf{W}}^\trans\mbf{X}_+\tilde{\mbf{V}}\tilde{\mbs{\Sigma}}^{-1}.
\end{equation}
Note that the eigenvalues of $\tilde{\mbf{A}}$ and $\mbf{A}$ are the same and that the eigenvectors of $\mbf{A}$ can be recovered as
\begin{equation}
    \mbs{\phi} = \mbf{X}_+\tilde{\mbf{V}}\tilde{\mbs{\Sigma}}^{-1}\mbf{q},
\end{equation}
where $\mbf{q}$ is an eigenvector of $\tilde{\mbf{A}}$. DMD is useful for studying the growth and decay rates, and the dynamic modes, of high-dimensional systems. Note that since DMD is used with \eqref{dynamics_eq}, which describes the dynamics of a linear system, DMD can be used in tandem with Koopman operator theory to represent nonlinear systems as approximate linear systems. Additionally, DMDc is a method presented in \cite{Proctor2016} to perform DMD on a dataset with inputs.

\section{Total DMD with inputs} \label{TDMD}

In practice, data collected from real systems is corrupted by noise that can lead to a bias in the identified model. DMD and the standard least-squares problem do not mitigate the effect of noise in the identification process of the Koopman matrix. In \cite{Proctor2016}, TDMD, which is a version of DMD inspired by total least-squares, is introduced to reduce the effect of noise by using a projection step before computing the Koopman matrix with least squares. However, TDMD is only used with datasets that consist of collected measurements and no inputs. 

In this section, the existing TDMD method is reviewed and then modified to accommodate inputs. The proposed method, TDMD with inputs (TEDMD), does not only reduce the bias found in the dynamics matrix, but also reduces the bias in the input matrix.

\subsection{Total least-squares formulation of DMD} \label{TDMD_van}

Consider a dynamical system with no inputs whose Koopman matrix is computed by solving the least-square problem
\begin{equation}
    \mbf{U} = \arg \min_{\mbf{U}^\star} \hspace{10pt} \left\|\mbs{\Theta}_+ - \mbf{U}^\star\mbs{\Theta}\right\|^2_\frob. \label{states_least}
\end{equation}
The solution to \eqref{states_least} is
\begin{equation} \label{states_soln}
    \mbf{U} = \mbs{\Theta}_+\mbs{\Theta}^\dagger, 
\end{equation}
where 
\begin{equation}
    \mbs{\Theta} = \begin{bmatrix}
        \mbs{\vartheta}_0 & \mbs{\vartheta}_1 & \cdots & \mbs{\vartheta}_{q-1} 
    \end{bmatrix} \in \mathbb{R}^{p_\vartheta \times q},
\end{equation}
and $\mbs{\Theta}_+$ is defined by \eqref{Theta_+}. The solution in \eqref{states_soln} to the least-squares problem in \eqref{states_least} is computed under the assumption that the noise corrupting the data is only present in $\mbs{\Theta}_+$. Specifically, the least-squares problem in \eqref{states_least} is equivalent to
\begin{align} \label{here1}
    &\min \hspace{10pt} J(\mbf{U}_\mathrm{ls}, \mbs{N}_{\mbs{\Theta}_+}) = \|\mbs{N}_{\mbs{\Theta}_+}\|_\frob \\
    &\text{s.t.} \hspace{10pt} \mbs{\Theta}_+ + \mbs{N}_{\mbs{\Theta}_+} = \mbf{U}_\mathrm{ls}\mbs{\Theta}, \label{here2}
\end{align}
 where $\mbs{N}_{\mbs{\Theta}_+}$ is the added noise to $\mbs{\Theta}_+$.
This one-sided assumption creates a bias in the identified Koopman matrix. Total least-squares addresses this problem by assuming that noise is present in both snapshot matrices, such that the problem in \eqref{here1} and \eqref{here2} becomes \cite{Hemati2017}
\begin{align}
    &\min \hspace{10pt} J(\mbf{U}_\mathrm{tls}, \mbs{N}_{\mbs{\Theta}}, \mbs{N}_{\mbs{\Theta}_+}) = \left\|\begin{bmatrix}
        \mbs{N}_{\mbs{\Theta}} \\
        \mbs{N}_{\mbs{\Theta}_+}
    \end{bmatrix}\right\|_\frob \label{TLS1}\\
    &\text{s.t.} \hspace{10pt} \mbs{\Theta}_+ + \mbs{N}_{\mbs{\Theta}_+} = \mbf{U}_\mathrm{tls}(\mbs{\Theta} + \mbs{N}_{\mbs{\Theta}}). \label{TLS2}
\end{align}
TDMD \cite{Hemati2017}, which is inspired by total least-squares, involves a projection step and an identification step \cite{Hemati2017}. Consider the augmented matrix
\begin{equation} \label{aug_1}
    \mbf{Z} = \begin{bmatrix}
        \mbs{\Theta} \\
        \mbs{\Theta}_+
    \end{bmatrix}.
\end{equation}  
As opposed to regular least squares, TDMD does not only project $\mbs{\Theta}$ and $\mbs{\Theta}_+$ on the range of $\mbs{\Theta}^\trans$, but on the range of $\mbf{Z}^\trans$, which captures both the range of $\mbs{\Theta}^\trans$ and $\mbs{\Theta}_+^\trans$.
Consider the SVD
\begin{equation} \label{svd1}
    \mbf{Z} = \mbf{W}\mbs{\Sigma}\mbf{V}^\trans,
\end{equation}
where $\mbf{W} \in \mathbb{R}^{(p_\vartheta + p_\vartheta) \times r}$, $\mbs{\Sigma}\in \mathbb{R}^{r \times r}$, $\mbf{V} \in \mathbb{R}^{q \times r}$. The transpose of $\mbf{Z}$ is given by 
\begin{equation}
    \mbf{Z}^\trans = \mbf{V}\mbs{\Sigma}\mbf{W}^\trans.
\end{equation}
Note that the SVD of either $\mbf{Z}$ or $\mbf{Z}^\trans$ can be truncated. Throughout this paper, it's assumed the SVD of $\mbf{Z}$ is truncated. The projection matrix onto the range of $\mbf{Z}^\trans$ is 
\begin{equation}
    \mathbb{P}_{\mbf{Z}^\trans} = \mbf{V}\mbf{V}^\trans.
\end{equation}
Then, let the snapshot matrices $\mbs{\Theta}$ and $\mbs{\Theta}_+$ be projected onto the range of $\mbf{Z}^\trans$, such that the approximate Koopman dynamics become
\begin{align}
    \mbs{\Theta}_+\mathbb{P}_{\mbf{Z}^\trans} &= \mbf{U}_\mathrm{tdmd}\mbs{\Theta}\mathbb{P}_{\mbf{Z}^\trans}, \label{keypoint} \\
    \bar{\mbs{\Theta}}_+ &= \mbf{U}_\mathrm{tdmd}\bar{\mbs{\Theta}}, \label{blablabla}
\end{align}
where $\bar{\mbs{\Theta}} = \mbs{\Theta}\mathbb{P}_{\mbf{Z}^\trans}$ and $\bar{\mbs{\Theta}}_+ = \mbs{\Theta}\mathbb{P}_{\mbf{Z}^\trans}$.
Several methods are presented in the literature \cite{Gavish2014, Vannieuwenhoven2012, Rust1998} to find the optimal singular value threshold at which to truncate the SVD in order to keep the main low-rank spatiotemporal features while removing less important features \cite{Kutz2016}. One method to obtain the optimal singular value threshold is the singular value hard threshold method introduced in \cite{Gavish2014}. In practice, for a shorter computing time, the snapshot matrices $\bar{\mbs{\Theta}} \in \mathbb{R}^{p_\vartheta \times q}$ and $\bar{\mbs{\Theta}}_+ \in \mathbb{R}^{p_\vartheta \times q}$ do not have to be explicitly used, since when the number of snapshots $q$ is high, then
solving for $\mbf{U}_\mathrm{tdmd}$ in \eqref{blablabla} becomes computationally demanding. Instead, the smaller matrices 
\begin{equation} \label{here3}
    \hat{\mbs{\Theta}} = \mbs{\Theta}\mbf{V} \in \mathbb{R}^{p_\vartheta \times r}, \hspace{10pt} \hat{\mbs{\Theta}}_+ = \mbs{\Theta}_+\mbf{V} \in \mathbb{R}^{p_\vartheta \times r}
\end{equation}
are computed to form \cite{Hemati2017} 
\begin{equation} \label{here4}
    \hat{\mbs{\Theta}}_+ = \mbf{U}_\mathrm{tdmd}\hat{\mbs{\Theta}}.
\end{equation}
After the projection step taken in \eqref{here3}, the Koopman matrix is computed by solving the least-squares problem in \eqref{here4}, which results in 
\begin{equation} \label{tdmd}
    \mbf{U}_\mathrm{tdmd} = \hat{\mbs{\Theta}}_+\hat{\mbs{\Theta}}^\dagger.
\end{equation}
In this section, TDMD is introduced as a method that computes a Koopman matrix with reduced bias by first projecting the snapshot matrices onto the range of an augmented matrix and then computing the Koopman matrix by solving the least-squares problem in \eqref{tdmd}.

\subsection{Modification of TDMD to include inputs} \label{sec: TDMDc}

TDMD is presented as a two-step method that reduces the bias in the dynamics matrix through a projection step and an identification step. However, TDMD is a method that reduces the impact of noise when the dataset used for identification has no inputs. Next, a modification to TDMD, where inputs are considered in the TDMD framework to identify the dynamics and input matrices with reduced bias is proposed.

For a system with inputs, its dynamics and input matrices can be computed using the least-squares solution
\begin{equation} \label{dmd_inputs}
    \begin{bmatrix}
        \mbf{A}_\mathrm{ls} & \mbf{B}_\mathrm{ls}
    \end{bmatrix} = \mbs{\Theta}_+\mbs{\Psi}^\dagger,
\end{equation}
where $\mbs{\Theta}_+$ and $\mbs{\Psi}$ are defined in \eqref{Theta_+} and \eqref{Psi}, respectively. The formulation of the problem in \eqref{dmd_inputs} also assumes that noise is present only in $\mbs{\Theta_+}$. Therefore, an equivalent method to TDMD must be derived to consider noise in both matrices. As in Section \ref{TDMD_van}, consider the augmented snapshot matrix
\begin{equation} \label{aug_2}
    \mbf{T} = \begin{bmatrix}
        \mbs{\Psi} \\
        \mbs{\Theta}_+
    \end{bmatrix},
\end{equation}
with SVD
\begin{equation}
    \mbf{T} = \mbf{W}\mbs{\Sigma}\mbf{V}^\trans.
\end{equation}
The projection matrix onto the range of $\mbf{T}^\trans$ is
\begin{equation} \label{here69}
    \mathbb{P}_{\mbf{T}^\trans} = \mbf{V}\mbf{V}^\trans.
\end{equation}
Then, let the snapshot matrices $\mbs{\Psi}$ and $\mbs{\Theta}_+$ be projected onto the range of $\mbf{T}^\trans$ as
\begin{align}
    \mbs{\Theta}_+\mathbb{P}_{\mbf{T}^\trans} &= \mbf{U}_\mathrm{tdmdc}\mbs{\Psi}\mathbb{P}_{\mbf{T}^\trans}, \label{keypoint_2} \\
    \bar{\mbs{\Theta}}_+ &= \mbf{U}_\mathrm{tdmdc}\bar{\mbs{\Psi}},
\end{align}
where $\bar{\mbs{\Psi}} = \mbs{\Psi}\mathbb{P}_{\mbf{T}^\trans}$ and $\bar{\mbs{\Theta}}_+ = \mbs{\Theta}_+\mathbb{P}_{\mbf{T}^\trans}$.

Finally, the Koopman matrix with reduced bias identified from a dataset with inputs is computed with the least-squares solution
\begin{equation} \label{tdmdc}
    \mbf{U}_\mathrm{tdmdc} = \hat{\mbs{\Theta}}_+\hat{\mbs{\Psi}}^\dagger,
\end{equation}
where 
\begin{equation} \label{projection}
    \hat{\mbs{\Psi}} = \mbs{\Psi}\mbf{V}, \hspace{20pt} \hat{\mbs{\Theta}}_+ = \mbs{\Theta}_+\mbf{V}.
\end{equation}
The derivation for the TEDMD follows in parallel that of the TDMD. The difference lies in the construction of the augmented matrices described by \eqref{aug_1} and \eqref{aug_2}. Although similar to TDMD, TEDMD allows for a much wider range of applications where noise corrupts data with inputs.

\section{Formulation of the optimization problem}

The Koopman operator approximation problem with TEDMD can be computed traditionally using \eqref{tdmdc} or equivalently as 
\begin{equation}
    \mbf{U} = \argmin_{\mbf{U}^\star} \hspace{10pt} \left\|\left(\hat{\mbs{\Theta}}_+ - \mbf{U}^\star\hat{\mbs{\Psi}}\right)\mbs{\Omega}\right\|^2_\frob, \label{tdmdc_least}
\end{equation}
where $\mbs{\Omega}$ is a weight nominally set to the identity matrix. As highlighted in \cite{Lortie2024, Dahdah2022}, the Koopman operator optimization problem described by \eqref{tdmdc_least} can be formulated in a modular fashion as a series of LMIs. This feature allows the problem to be written in tandem with LMI constraints to enforce specific requirements on the Koopman matrix. For models of real systems, one potential requirement is asymptotic stability. In particular, any real system that is asymptotically stable must have a Koopman representation that is also asymptotically stable.

For a discrete-time system to be asymptotically stable, its dynamics matrix must have its eigenvalues bounded strictly within the unit circle \cite{Ogata1995}. To enforce asymptotic stability on the Koopman system with reduced bias, the BMI constraint \cite{Ogata1995, Dahdah2022}
\begin{equation}
    \begin{bmatrix}
        \bar{\rho}\mbf{P} & \mbf{A}\mbf{P} \\
        \star &
        \bar{\rho}\mbf{P}
    \end{bmatrix} > 0,
\end{equation}
with 
\begin{equation}
    \mbf{P} > 0,
\end{equation}
where $\star$ defines the transpose of its off-diagonal counterpart and $\bar{\rho}$ is the spectral bound on the eigenvalues of $\mbf{A}$, is added to the optimization problem in \eqref{tdmdc_least}. In particular, the optimization problem is now
\begin{align}
   & \mbf{U} = \argmin_{\mbf{A}, \mbf{B}, \mbf{P}} \hspace{10pt} \left\|\left(\hat{\mbs{\Theta}}_+ - \begin{bmatrix}
        \mbf{A} & \mbf{B}
    \end{bmatrix}\hat{\mbs{\Psi}}\right)\mbs{\Omega}\right\|^2_\frob \label{cost_tdmdc}\\
    & \text{s.t.} \hspace{10pt} \mbf{P} > 0, \hspace{10pt}
    \begin{bmatrix}
        \bar{\rho}\mbf{P} & \mbf{A}\mbf{P} \\
        \star &
        \bar{\rho}\mbf{P}
    \end{bmatrix} > 0. \label{here40}
\end{align}
Note that the optimization problem described in \eqref{cost_tdmdc} and \eqref{here40} is not convex because of the bilinear term $\mbf{A}\mbf{P}$. To transform the above nonconvex problem as a convex problem, the change of variable $\mbf{F} = \mbf{A}\mbf{P}$ proposed in \cite{Mabrok2023, Lortie2024, Hara2020, Hara2021} is used. Setting the weight $\mbs{\Omega}$ equal to 
\begin{equation}
    \mbs{\Omega} = \hat{\mbs{\Psi}}^\trans (\hat{\mbs{\Psi}} \hat{\mbs{\Psi}}^\trans)^\dagger \bar{\mbf{P}}, \label{p_const}
\end{equation}
in \eqref{cost_tdmdc} yields the convex optimization problem
\begin{align}
   & \min \hspace{10pt} J(\mbf{F}, \mbf{B}, \mbf{P}) = \left\|\hat{\mbf{G}} \hat{\mbf{H}}^\dagger \bar{\mbf{P}} - \begin{bmatrix}
        \mbf{F} & \mbf{B}
    \end{bmatrix}\right\|^2_\frob \label{cost_1} \\
    & \text{s.t.} \hspace{10pt} \mbf{P} > \epsilon, \hspace{10pt}
    \begin{bmatrix}
        \bar{\rho}\mbf{P} & \mbf{F} \\
        \star &
        \bar{\rho}\mbf{P}
    \end{bmatrix} > 0, \label{here41}
\end{align}
where 
\begin{equation}
    \bar{\mbf{P}} = \begin{bmatrix} 
        \mbf{P} & \textbf{0} \\ 
        \textbf{0} & \textbf{1}
    \end{bmatrix},
\end{equation}
$\hat{\mbf{G}} = \hat{\mbs{\Theta}}_+\hat{\mbs{\Psi}}^\trans$, $\hat{\mbf{H}} = \hat{\mbs{\Psi}}\hat{\mbs{\Psi}}^\trans$, $\mbf{F} = \mbf{A}\mbf{P}$, and $\epsilon$ is added to set a lower bound on $\mbf{P}$ in order to not let $\mbf{P}$ be prioritize when minimizing the cost function in \eqref{cost_1}. Then, a slack variable is introduced to decompose the cost function in \eqref{cost_1} as a combination of LMI constraints~\cite{Caverly2019, Dahdah2022, Lortie2024}, such that the final optimization becomes 
\begin{align}
& \min \hspace{10pt} J(\gamma, \mbf{F}, \mbf{B}, \mbf{P}, \hat{\mbf{R}}) = \gamma \label{cost_final} \\
    & \textrm{s.t.} \hspace{10pt} \trace{(\hat{\mbf{R}})} < 1, \hspace{10pt}
    \hat{\mbf{R}} > 0, \hspace{10pt}
    \begin{bmatrix}
        \hat{\mbf{R}} & \left(\hat{\mbf{G}} \hat{\mbf{H}}^\dagger \bar{\mbf{P}} - \begin{bmatrix}
            \mbf{F} & \mbf{B}
        \end{bmatrix}\right)^\trans \\ \star & \gamma \textbf{1}
    \end{bmatrix} > 0, \label{here42.2} \\
    & \mbf{P} > \epsilon, \hspace{10pt}
    \begin{bmatrix}
        \bar{\rho}\mbf{P} & \mbf{F} \\
        \star &
        \bar{\rho}\mbf{P}
    \end{bmatrix} > 0, \label{here42}
\end{align}
where $\mbf{A}$ can be solved for using $\mbf{A} = \mbf{F}\mbf{P}^{-1}$. The Koopman matrix is then computed by solving the optimization problem in \eqref{cost_final}--\eqref{here42} where
\begin{equation}
    \mbf{U} = \begin{bmatrix}
        \mbf{A} & \mbf{B}
    \end{bmatrix}.
\end{equation}
A detailed derivation from \eqref{tdmdc_least} to \eqref{here42} is presented in \cite{Lortie2024}.

In this section, the Koopman operator approximation problem is posed as a convex optimization problem with LMI constraints that enforce asymptotic stability on the system. Enforcing asymptotic stability is important, since the approximate Koopman representation of the asymptotically stable system must be properly representative.

\section{Results and discussion} \label{results_section}

To validate that the proposed method, TDMD with inputs and asymptotic stability constraint (TEDMD-AS), solves for an asymptotically stable approximate Koopman system with reduced bias and, as will be shown, outperforms other state-of-the-art methods when testing using a simulated Duffing oscillator dataset and an experimental soft robot arm dataset. In this section, TEDMD-AS is compared to total DMD with inputs (TEDMD), EDMD, EDMD with an asymptotic stability constraint (EDMD-AS), forward-backward EDMD with inputs (fbEDMD), and forward-backward EDMD with inputs and asymptotic stability constraints (fbEDMD-AS) to assess its prediction and bias reduction performance. The methods using asymptotic stability are also compared to the methods that do not use constraints. The Koopman matrices identified from TEDMD and TEDMD-AS are computed using the optimization problem described in \eqref{tdmdc_least} and \eqref{cost_final}--\eqref{here42}, respectively. EDMD-AS and fbEDMD-AS compute Koopman matrices using the optimization problems described in \cite[\S 5]{Lortie2024}. As for EDMD, the Koopman matrix is generated by \texttt{pykoop} \cite{Pykoop2021}, an open-source Koopman operator identification library. Forward-backward EDMD uses \texttt{pykoop} and the method described in \cite[\S 3.2]{Lortie2024}. All three methods with asymptotic stability constraints use the spectral radius bound $\bar{\rho} = 0.99999$ \cite{Lortie2024}.

\subsection{Simulation results} \label{simulated_results}

A simulated dataset of a Duffing oscillator is used to compare approximate Koopman systems generated by the different methods. In particular, the system that generates the dataset for this section is described by the ordinary differential equation
\begin{equation} \label{Duff}
    m\ddot{x}(t)+ c\dot{x}(t) + k_1x(t) + k_2x(t)^3 = f(t),
\end{equation}
where the mass $m = 0.1$ kg, damping coefficient $c = 0.01$ Ns/m, linear spring constant $k_1 = 0.1$ N/m and nonlinear spring constant $k_2 = 0.001$ $\text{N/m}^3$.  To generate the discretized dataset, Euler's forward method \cite[\S 6.1]{Sauer2017} is used. Additionally, this simulated dataset consists of 20 training episodes and 2 testing episodes. To lift the simulated data into the lifted space, the chosen lifting functions are all the different combinations of second-order monomials and ten radial basis functions of the form 
\begin{equation} \label{lift_func}
    \psi^\mathrm{rbf}_i(\mbf{x}) = r_i^2\ln\left(r_i\right), \;\;\;\; i = 1, \dots, 10,
\end{equation}
where 
\begin{equation}
    r_i = \alpha\left\|\mbs{\psi}^\mathrm{poly}(\mbf{x}) - \mbf{c}_i\right\| + \delta,
\end{equation}
$\mbs{\psi}^\mathrm{poly}(\mbf{x})$ is a vector-valued lifting function made of all second-order monomial combinations, $\mbf{c}_i$ is a center generated by a Latin hypercube sampling algorithm~\cite{Mckay2000, Eglajs1977}, $\alpha = 0.1$ is a parameter used to define the scaling, and $\delta = 0.001$ ensures that the computed number is defined. Note that the values for $\alpha$ and $\delta$ are tuned based on the dataset and chosen lifting functions to obtain the best fit. Additionally, the signal-to-noise ratio (SNR), computed as
\begin{equation}
    \text{SNR} = 10\log\left(\frac{\sigma_x^2}{\sigma_n^2}\right),
\end{equation}
where $\sigma_x$ is the standard deviation of the original signal and $\sigma_n$ is the standard deviation of the added noise, is used to describe the amount of noise added to the original trajectory of the system. The added noise $\mbf{v}_k$ found in the measurement equation described by \eqref{output_eq} is simulated by white noise of distribution $\mbf{v}_k \sim \mathcal{N}(\mbf{0}, \mbs{\Sigma}_{\mathrm{v}})$, where the covariance matrix $\mbs{\Sigma}_{\mathrm{v}}$ is set to obtain a particular SNR. In this paper, SNRs of 18 dB and 28 dB are evaluated, where 18 dB corresponds to more noise than 28 dB.

Before diving into the results of TEDMD, a particular challenge in the projection step of TEDMD must be discussed. When the left singular vectors of the augmented matrix are used to project the snapshot matrices in \eqref{keypoint} of TEDMD, the condition number of both snapshot matrices $\mbs{\Theta}_+$ and $\mbs{\Psi}$ increases, making the least-squares problem in \eqref{tdmdc_least} have larger condition numbers than the regular least-squares problem in \eqref{vanilla_least} \cite{Markovsky2007}. A solution to this problem is to truncate the SVD problem as mentioned in Section \ref{sec: TDMDc}. There exists a trade-off between removing too few or too many singular values to form $\mbf{V}$ described in \eqref{here69}. When truncating many singular values, the condition number of the projected matrices decreases, but some information contained in the original snapshot matrices is lost. On the other hand, keeping too many singular values leads to the solver finding no solution due to large condition numbers of the projected snapshot matrices. The optimal truncation value for $\mbf{V}$ must therefore be large enough to keep all the important singular vectors, while maintaining a low enough condition number to allow for a solution of the least-squares problem.  Although the literature provides different methods, such as the hard threshold method \cite{Gavish2014} to select the optimal singular value at which truncation should occur, it was noticed that none of these methods would consistently generate the optimal truncation value for all the range of SNRs considered in this paper. Therefore, for all considered SNRs, the number of singular values to keep is individually hand-picked as the highest integer possible, while having a low enough condition number on both projected snapshot matrices to allow for a solution in the projected least-squares problem. 

\begin{figure}
    \centering
    \includegraphics[scale=1.1]{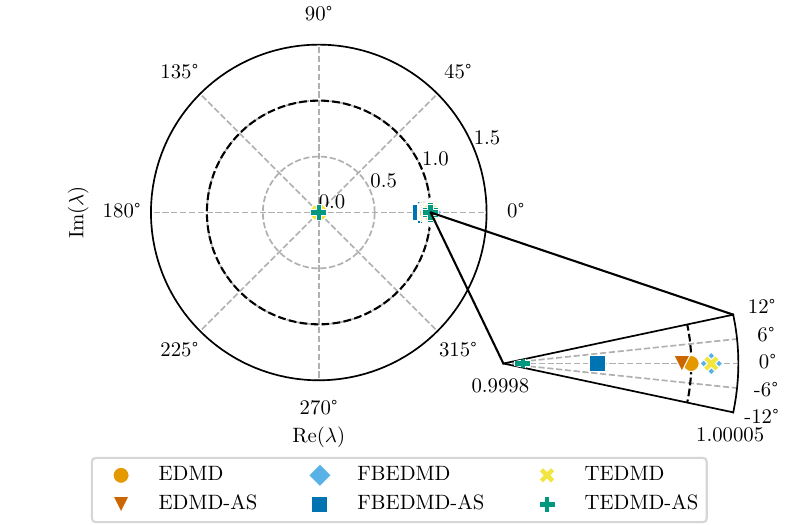}
    \caption{Eigenvalues of the identified Koopman matrices using the simulated Duffing oscillator dataset at an SNR of 28 dB. All the methods with asymptotic stability constraints, EDMD-AS, fbEDMD-AS, and TEDMD-AS, have their largest eigenvalue bounded strictly within the unit circle, which means that the identified Koopman systems are asymptotically stable. Forward-backward EDMD and TEDMD produce unstable systems with some eigenvalues of the Koopman matrices outside the unit circle. Although EDMD does not enforce asymptotic stability, it still identified a stable Koopman system.}
    \label{duff_2}
\end{figure}

The ability of a method to enforce asymptotic stability on the computed Koopman system is assessed by observing if all of the eigenvalues of the system's dynamics matrix are within the unit circle of the complex plane. Note that the ability of a method with no asymptotic stability constraint to generate an asymptotic stability approximate Koopman system depends on many factors such as the dataset and the chosen lifting functions. In particular, just as EDMD-AS, fbEDMD-AS, and TEDMD-AS compute dynamics matrices with eigenvalues within the unit circle in Figure \ref{duff_2}, EDMD also computes a dynamics matrix with all of its eigenvalues inside the unit circle. Consequently, for simplicity, when comparing trajectories generated from different identification methods, the trajectories of EDMD, fbEDMD and TEDMD are omitted as they can generate unstable trajectories, as shown from the eigenvalues in Figure \ref{duff_2}.

\begin{figure}
\renewcommand\thesubfigure{\roman{subfigure}}
    \centering
    \begin{subfigure}{\linewidth}
    \includegraphics[width=\linewidth]{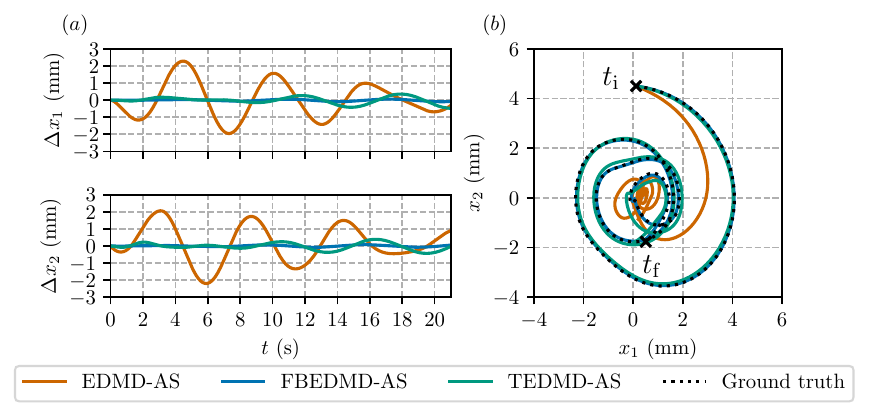}
    \caption{Prediction error plot $(a)$ and multi-step trajectory $(b)$ for Koopman matrices identified with an SNR of 28 dB. The trajectory generated by the Koopman matrix identified with EDMD-AS exhibits a large bias compared to fbEDMD-AS and TEDMD-AS, which is reflected by larger decay rates. While both fbEDMD-AS and TEDMD-AS identify Koopman matrices that predict the trajectory with a low error, fbEDMD-AS shows a lower error than TEDMD-AS.}
    \label{duff_0}
    \end{subfigure}
    \vfill
    \begin{subfigure}{\linewidth}
        \includegraphics[width=\linewidth]{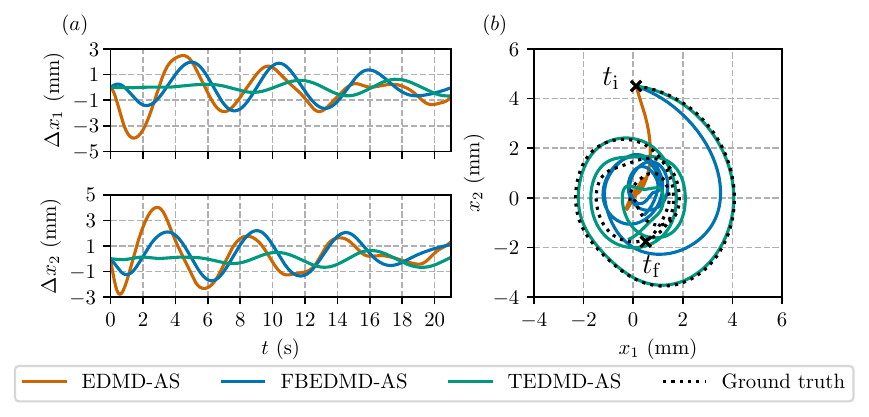}
    \caption{Prediction error plot $(a)$ and multi-step trajectory $(b)$ for Koopman matrices identified with an SNR of 28 dB. The trajectory generated by the Koopman matrix identified with EDMD-AS exhibits a large bias compared to fbEDMD-AS and TEDMD-AS, which is reflected by larger decay rates. Compared to TEDMD-AS, fbEDMD-AS identifies a Koopman matrix with a higher bias. The Koopman matrix identified with TEDMD-AS predicts the trajectory with the lowest error.}
    \label{duff_1}
    \end{subfigure}
    \caption{Prediction error plots and multi-step trajectories of the first test episode for all three asymptotically stable Koopman systems identified with the simulated Duffing oscillator dataset at an SNR of (i) 28 dB and (ii) 18 dB. At each step of the prediction, the states are recovered and re-lifted. The prediction starts at $t_\mathrm{i} = 0 \;\mathrm{s}$ and finishes at $t_\mathrm{f} = 21 \; \mathrm{s}$.}
\end{figure}

A useful feature of dynamic models is to predict a system's response to a given set of initial conditions. Next, the approximate Koopman models computed by the different methods are used to compare trajectory prediction errors, where the smallest error corresponds to the approximate Koopman model that generated the closest trajectory to the true trajectory. To emphasize the ability of TEDMD-AS and fbEDMD-AS to reduce the bias in the system's trajectory in the presence of noise, Figure \ref{duff_0} and Figure \ref{duff_1} show generated trajectories when the dataset used for identification has an SNR of 28 dB and 18 dB, respectively. The trajectory generated by the Koopman matrix computed with EDMD-AS in Figure \ref{duff_0} is shown to have a significant bias compared to the trajectories obtained with TEDMD-AS and fbEDMD-AS. In this case, the bias in the Koopman matrix is analogous to a bias in the eigenvalues, leading to a higher decay rate. When comparing fbEDMD-AS and TEDMD-AS at an SNR of 28 dB in Figure \ref{duff_0}, fbEDMD-AS has a lower error than TEDMD-AS. However, both fbEDMD-AS and TEDMD-AS have a reduced bias in their trajectories compared to EDMD-AS. When the dataset is corrupted by noise with an SNR of 18 dB, the trajectory generated with TEDMD-AS has a lower error than both fbEDMD-AS and EDMD-AS. It can therefore be said that above a certain threshold of noise, TEDMD-AS outperforms both EDMD-AS and fbEDMD-AS.

\subsection{Experimental results}

In this section, the different identification methods are tested with an experimental dataset of a soft robot arm~\cite{Bruder2019, Bruder2020}. Although the soft robot arm has inputs, methods such as DMDc, EDMD, fbEDMD, and the one presented in this paper can still be used to identify an approximation of the Koopman operator with a dataset that includes state and input measurements. Assuming that the dynamics of the system are time-independent \cite{Brunton2021} is a common practice in the field, and is validated with convincing results \cite{Williams2015, Proctor2016, Dahdah2022, Kaiser2021}. This soft robot arm is controlled by three pressure regulators, which act as this system's control inputs. The laser pointer mounted on the end effector of this soft robot points onto a board, where the $x$-$y$ coordinates of the point represent this system's states and output measurement. The provided dataset consists of 13 training training episodes and 4 test episodes. Although all experimental datasets have some noise due to the noisy nature of sensors, additional white noise modeled as $\mbf{w}_k \sim \mathcal{N}(\mbf{0}, \mbs{\Sigma}_{\mathrm{w}})$ is added to the measurement equation in \eqref{output_eq} to emphasize the ability to reduce the bias in the Koopman system due to noise for TEDMD methods. Similarly to the simulated dataset, the covariance matrix $\mbs{\Sigma}_{\mathrm{w}}$ is set so that the experimental dataset with added noise has a specific SNR. In this section, the concept of \emph{true} Koopman matrix is discussed. The true Koopman matrix refers to the Koopman matrix identified with the original dataset without added noise as opposed to the predicted Koopman matrix, which is the Koopman matrix identified with the added noise. The lifting functions used to fit the Koopman model to the data are the same as described in Section \ref{simulated_results}, but with a shape factor $\alpha = 0.5$.

\begin{figure}
    \centering
    \includegraphics[scale=1.1]{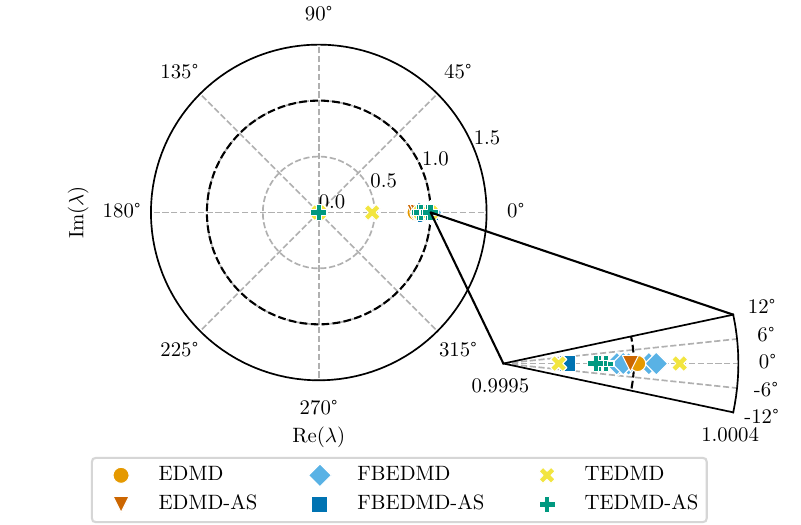}
    \caption{Eigenvalues of the identified Koopman matrices using the experimental soft robot arm dataset at an SNR of 28 dB. All the methods with asymptotic stability constraints, EDMD-AS, fbEDMD-AS, and TEDMD-AS, have their largest eigenvalue bounded strictly within the unit circle, which means that the identified Koopman systems are asymptotically stable. Extended DMD, fbEDMD, and TEDMD produce unstable systems with some eigenvalues of the Koopman matrices outside the unit circle.}
    \label{soft_0}
\end{figure}

The effect of adding asymptotic stability constraints to the approximate Koopman identification problem is demonstrated in Figure \ref{soft_0}, where EDMD, fbEDMD, and TEDMD have eigenvalues located outside the unit circle. Consequently, Koopman systems obtained with EDMD, fbEDMD, and TEDMD are unstable and therefore will not be used for trajectory prediction, since the predictions will eventually diverge to infinity. When asymptotic stability constraints are used, as with EDMD-AS, fbEDMD-AS, and TEDMD-AS, all eigenvalues of the computed approximate Koopman matrices are strictly within the unit circle. Note that EDMD identified a Koopman matrix with eigenvalues outside the unit circle in Figure \ref{soft_0}, but not in Figure \ref{duff_2}, which shows that asymptotic stability constraints are important in order to have all eigenvalues within the unit circle.

\begin{figure}[H]
\renewcommand\thesubfigure{\roman{subfigure}}
     \centering
     \begin{subfigure}{\linewidth}
        \includegraphics[height=0.45\linewidth]{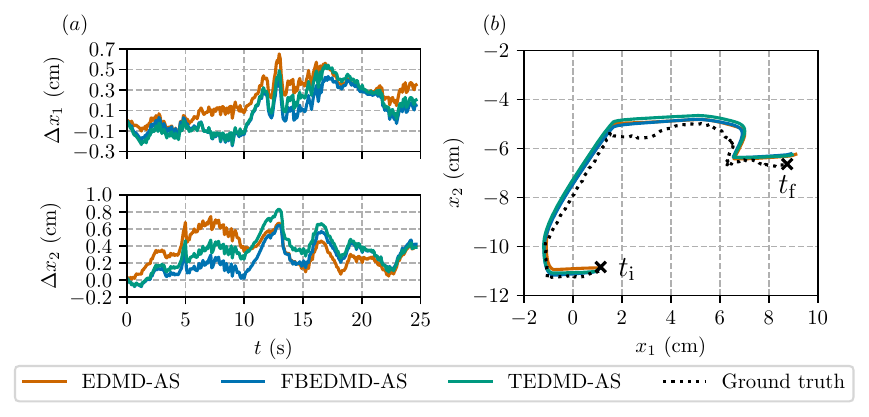}
        \caption{Prediction error plot $(a)$ and multi-step trajectory $(b)$ for Koopman matrices identified with an SNR of 28 dB. All three Koopman matrices identified with EDMD-AS, fbEDMD-AS, and TEDMD-AS predict trajectories with similar errors.}
        \label{soft_1}
        \end{subfigure}
        \vfill
        \begin{subfigure}{\linewidth}
        \includegraphics[height=0.45\linewidth]{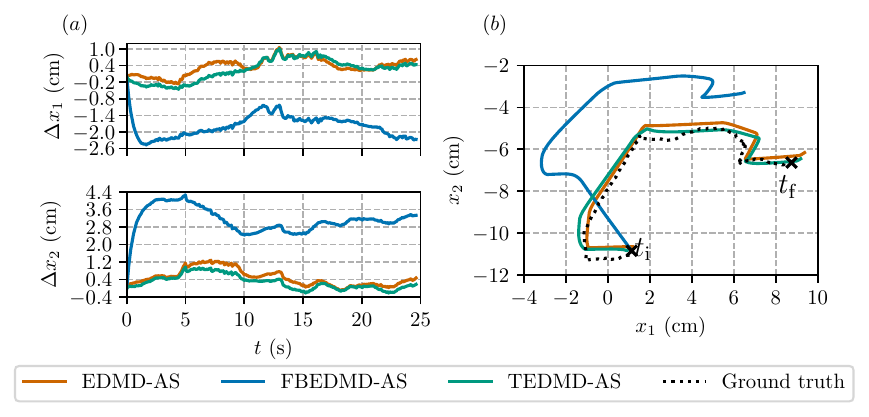}
        \caption{Prediction error plot $(a)$ and multi-step trajectory $(b)$ for Koopman matrices identified with an SNR of 18 dB. The trajectory generated by the Koopman matrix identified with fbEDMD-AS exhibits a large bias compared to EDMD-AS and TEDMD-AS. Both Koopman matrices computed with EDMD-AS and TEDMD-AS predict trajectories with similar errors.}
        \label{soft_3}
        \end{subfigure}
        \caption{Prediction error plots and multi-step trajectories of the second test episode for all three asymptotically stable Koopman systems identified with the experimental soft robot arm dataset at an SNR of (i) 28 dB and (ii) 18 dB. At each step of the prediction, the states are recovered and re-lifted. The prediction starts at $t_\mathrm{i} = 0 \;\mathrm{s}$ and finishes at $t_\mathrm{f} = 24.5 \; \mathrm{s}$.}
        \label{soft_traj}
\end{figure}

\begin{figure}[H]
\renewcommand\thesubfigure{\roman{subfigure}}
     \centering
     \begin{subfigure}{\linewidth}
        \includegraphics[width=\linewidth]{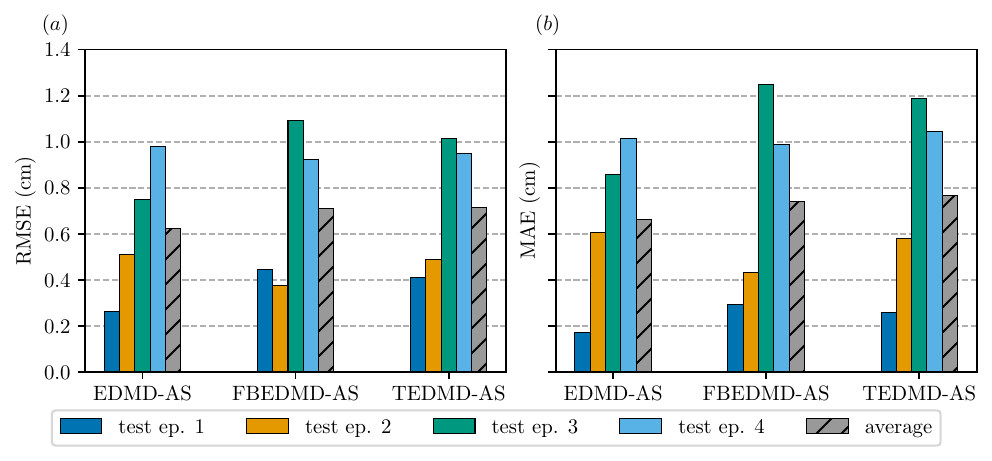}
        \caption{The root-mean-square (RMS) $(a)$ and mean $(b)$ multi-step prediction errors for Koopman matrices identified with an SNR of 18 dB. The prediction obtained with TEDMD-AS has higher averaged RMS and mean errors than both EDMD-AS and fbEDMD-AS. EDMD-AS has a higher mean error, but a lower RMS error than fbEDMD-AS.}
        \label{soft_2}
        \end{subfigure}
        \begin{subfigure}{\linewidth}
        \includegraphics[width=\linewidth]{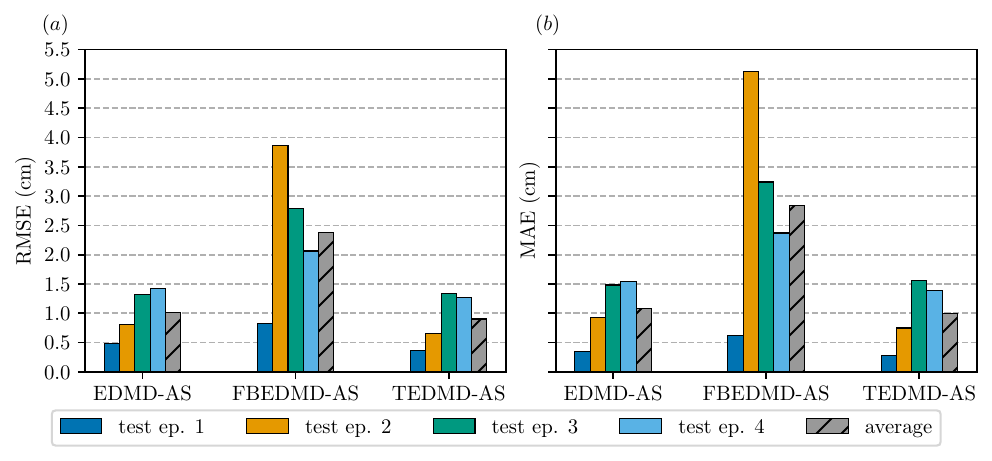}
        \caption{The root-mean-square (RMS) $(a)$ and mean $(b)$ multi-step prediction errors for Koopman matrices identified with an SNR of 18 dB. The prediction obtained with TEDMD-AS has lower averaged RMS and mean errors than both EDMD-AS and fbEDMD-AS. Forward-backward EDMD-AS has higher RMS and mean errors than both EDMD-AS and TEDMD-AS.}
        \label{soft_4}
        \end{subfigure}
        \caption{The root-mean-square (RMS) and mean multi-step prediction errors for Koopman matrices identified using the four test episodes given in the experimental soft robot arm dataset at an SNR of 28 dB (i) and 18 dB (ii).}
        \label{block_dia}
\end{figure}

When predicting trajectories of the soft robot arm, as shown in Figure \ref{soft_1}, there is no clear best method, since EDMD-AS, fbEDMD-AS, and TEDMD-AS all have similar errors. Therefore, to compare prediction performances, the mean absolute error (MAE), and root-mean-square error (RMSE) are used as metrics, where \cite{Qi2020}
\begin{equation}
    \text{MAE} = \frac{\sum^q_{k=1} \|\mbf{x}_k - \hat{\mbf{x}}_k\|_1}{q},
\end{equation}
\begin{equation}
    \text{RMSE} = \sqrt{\frac{\sum^q_{k=1}\|\mbf{x}_k - \hat{\mbf{x}}_k\|_2^2}{q}},
\end{equation}
$\mbf{x}_k$ is the ground truth of the trajectory at time step \emph{k}, and $\hat{\mbf{x}}_k$ is the predicted trajectory at time step \emph{k}. The RMSE is chosen as an error metric because it measures the dispersion of the predicted trajectory around the ground truth, while MAE is picked as another error metric since it represents the bias in the error~\cite{Karunasingha2022}. In Figure \ref{soft_2}, when the dataset has an SNR of 28 dB, TEDMD-AS has a higher averaged root-mean-square and mean error than both EDMD-AS and fbEDMD-AS. As for fbEDMD-AS, it generates a trajectory with a higher root-mean-square error, but lower mean error than EDMD-AS. From the results, TEDMD-AS does not perform better than EDMD-AS or fbEDMD-AS in terms of predictions when the soft robot arm dataset has an SNR of 28 dB. 

The same experiment can be done when a greater amount of noise is added to the dataset. With an SNR of 18 dB, Figure \ref{soft_3} shows that fbEDMD-AS does not perform well compared to EDMD-AS and TEDMD-AS. As explained in \cite{Lortie2024}, when too much noise is present in the dataset, the high condition numbers of the snapshot matrices cause the relationships between the forward- and backward-in-time dynamics to hold approximately instead of exactly, leading to a poor performance of fbEDMD-AS. Again, the root-mean-square and mean errors are used, since the predicted trajectory of EDMD-AS and TEDMD-AS are too similar to decide which method performs better. In Figure \ref{soft_4}, the predicted trajectory generated using TEDMD-AS has both a lower root-mean-square and mean error than EDMD-AS and fbEDMD-AS. From the results, TEDMD-AS performs better than both EDMD-AS and fbEDMD-AS in terms of trajectory prediction above a certain threshold of noise. As observed in Figure \ref{block_dia}, the trajectory predicted with a Koopman matrix identified using TEDMD-AS has lower RMS and mean errors when more noise is present in the dataset. Note that this result is similar to the one obtained with the simulated Duffing oscillator dataset. 

\begin{figure}
    \centering
    \includegraphics[scale=1.1]{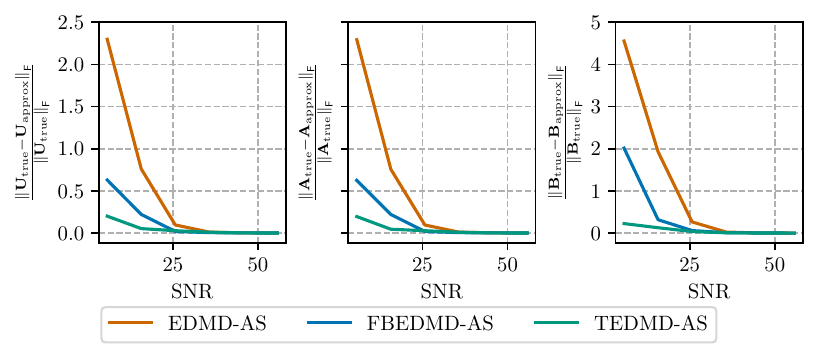}
    \caption{Relative error, using the Frobenius norm, between the approximated and true Koopman matrices, dynamics matrices, and input matrices at different SNRs. For all matrices computed with EDMD-AS, the relative error is greater than fbEDMD-AS and TEDMD-AS at all SNRs. Around an SNR of 25 dB, there is a threshold of added noise reached, where matrices computed with TEDMD-AS have a lower relative error than with fbEDMD-AS when more noise is added. At higher SNRs, fbEDMD-AS computes matrices with the lowest relative errors.}
    \label{soft_5}
\end{figure}

Considering that above a certain threshold of noise, TEDMD-AS identifies a Koopman matrix that predicts trajectories better than EDMD-AS and fbEDMD-AS with both the simulated and experimental datasets, it follows that the Koopman representation obtained with TEDMD-AS is more accurate than with EDMD-AS and fbEDMD-AS above a certain SNR. Note that for this section, the notion of accuracy is presented as a measure of proximity between the approximated Koopman matrix to the true Koopman matrix in terms of the Frobenius norm of the relative error. The approximated Koopman matrix is the Koopman matrix identified using the dataset with added noise and the true Koopman matrix is the Koopman matrix identified using the original dataset. In Figure \ref{soft_5}, the approximated Koopman matrix identified with TEDMD-AS is more accurate than with EDMD-AS and fbEDMD-AS around an SNR of 25 dB and below. Above an SNR of approximately 25 dB, although the identified Koopman matrices with all methods share a similar accuracy, fbEDMD-AS is more accurate than both EDMD-AS and TEDMD-AS. The same result is not only observed in the dynamics matrix of Figure \ref{soft_5}, but also in the input matrix, which indicates that the proposed adaptation of TDMD to TEDMD in Section \ref{sec: TDMDc} is valid. 

From the results of this section, EDMD-AS is shown to be the best method for predicting trajectories when a low amount of noise is present in the dataset. However, TEDMD-AS is the preferred method when a higher amount of noise is present in the dataset. In terms of accuracy, when accuracy is defined as the Frobenius norm of the relative error between the approximated and true Koopman matrices, EDMD-AS always gives a less accurate Koopman matrix over all SNRs, while TEDMD-AS is the best method at low SNRs and fbEDMD-AS at high SNRs.

\section{Conclusion} \label{conclusion_section}

Obtaining a Koopman representation of a real system with data-driven methods in the presence of noise is a difficult task, since a biased model can result. Moreover, regardless of the lifting functions used, it's important that the approximate Koopman model be asymptotically stable when the underlying dynamical system is asymptotically stable. The method proposed by this paper, TDMD with inputs and an asymptotic stability constraint, identifies an asymptotically stable approximate Koopman system with reduced bias when noisy data is used in the identification process, regardless of the lifting functions chosen. Using a simulated dataset of a Duffing oscillator and an experimental dataset of a soft robot arm, the proposed method is shown to compute a Koopman matrix that is closer to the noiseless solution and predicts a trajectory with a lower error than the state-of-the-art methods above a certain threshold of noise.

One limitation of this work is the requirement of truncation in the projection step of the proposed method in order to decrease the condition number. Future work will be focused on using regularization methods to better condition the least-squares problem instead of truncating singular values. The work developed in \cite{Dahdah2022}, which proposes the use of the $\mathcal{H}_\infty$ norm as a regularizer to improve the conditioning of the Koopman approximation problem, could potentially be adapted to the TEDMD framework.

\section{Funding} This work was supported by CIFAR, the Natural Sciences and Engineering Research Council of Canada (NSERC) discovery grants program, the Fonds de recherche du Québec – Nature et technologies (FRQNT), the Toyota Research Institute, the National Science Foundation Career Award (grant no.
1751093) and the Office of Naval Research (grant no. N00014-18-1-2575).

\section{Data accessibility}
All plots and datasets are accessible through the repository available at \url{https://github.com/decargroup/total_extended_dmd_koopman.git}.

\section{Acknowledgements} The authors thank Steven Dahdah for providing meaningful insight that helped push this research project forward. The authors thank Daniel Bruder, Xun Fu and Ram Vasudevan for graciously providing
the soft robot dataset used in this research.

\bibliographystyle{unsrt}
\bibliography{tedmd}

\end{document}